\documentclass[aps, prd, reprint, superscriptaddress]{revtex4-2}
\usepackage{graphicx}
\usepackage{bm}
\usepackage{amsmath}
\usepackage{amsfonts}
\usepackage{amssymb}
\usepackage[utf8]{inputenc}
\newcommand {\be} {\begin{equation}}
\newcommand {\ba} {\begin{eqnarray}}
\newcommand {\ee} {\end{equation}}
\newcommand {\ea} {\end{eqnarray}}

\begin{document}

\title{Elastic pion-proton and pion-pion scattering at high energies in holographic QCD}

\author{Zhibo Liu}
\email{202107020021014@ctgu.edu.cn}
\affiliation{College of Science, China Three Gorges University, Yichang 443002, People's Republic of China}
\affiliation{Center for Astronomy and Space Sciences, China Three Gorges University, Yichang 443002, People's Republic of China}
\author{Wei Xie}
\email{xiewei@ctgu.edu.cn}
\affiliation{College of Science, China Three Gorges University, Yichang 443002, People's Republic of China}
\affiliation{Center for Astronomy and Space Sciences, China Three Gorges University, Yichang 443002, People's Republic of China}
\author{Fei Sun}
\email{sunfei@ctgu.edu.cn}
\affiliation{College of Science, China Three Gorges University, Yichang 443002, People's Republic of China}
\affiliation{Center for Astronomy and Space Sciences, China Three Gorges University, Yichang 443002, People's Republic of China}
\author{Shuang Li}
\email{lish@ctgu.edu.cn}
\affiliation{College of Science, China Three Gorges University, Yichang 443002, People's Republic of China}
\affiliation{Center for Astronomy and Space Sciences, China Three Gorges University, Yichang 443002, People's Republic of China}
\author{Akira Watanabe}
\email{watanabe@ctgu.edu.cn (Corresponding author)}
\affiliation{College of Science, China Three Gorges University, Yichang 443002, People's Republic of China}
\affiliation{Center for Astronomy and Space Sciences, China Three Gorges University, Yichang 443002, People's Republic of China}

\date{\today}

\begin{abstract}
The total and differential cross sections of high energy pion-proton and pion-pion scattering are investigated in a holographic QCD model, focusing on the Regge regime in which the Pomeron exchange gives the dominant contribution to those cross sections.
The Pomeron is described by the Reggeized spin-2 particle propagator, and the hadron-Pomeron couplings are given by gravitational form factors of the hadrons, which are obtained from the bottom-up AdS/QCD models.
Since the parameters, which characterize the Pomeron trajectory, have been determined in a preceding study of the proton-proton scattering, the pion-proton cross sections can be expressed with a single adjustable parameter.
Once this parameter is determined by the experimental data of the pion-proton total cross section, its differential cross section and the pion-pion cross sections can be calculated without any additional parameter.
Although the currently available data are limited, it is found that our calculation is consistent with those.
Our predictions are explicitly shown, and the present model can be tested at the future experimental facilities.
\end{abstract}

\maketitle

\section{\label{sec:introduction}Introduction}
High energy hadron-hadron scattering has been one of the most important research topics in high energy physics over several decades, since its cross sections reflect the internal structure of the involved hadrons.
In particular, the proton-proton scattering has been intensively studied, and for instance, the recent experimental data for the high energy elastic scattering measured by the TOTEM collaboration at the LHC~\cite{TOTEM:2013lle,TOTEM:2016lxj,TOTEM:2017asr,TOTEM:2017sdy,TOTEM:2018hki,TOTEM:2018psk} have provided us with the valuable opportunities to deepen our understandings about the proton structure.
It is natural to have interest in other hadron-hadron scattering processes which involve other hadrons.
In this work, we focus on the pion involved processes, the elastic pion-proton and pion-pion scattering in the Regge regime.
Various preceding studies have been presented in Refs.~\cite{Pelaez:2003ky,Caprini:2011ky,Halzen:2011xc,Mathieu:2015gxa,Khoze:2017bgh}.
The pion is the lightest Nambu-Goldstone boson, and it is important to understand its structure to improve our knowledge of quantum chromodynamics (QCD).
We investigate the total and differential cross sections, and discuss the differences between those processes and the proton-proton case.

Elastic hadron-hadron scattering is a simple two-body process, and its cross sections shall be described by QCD.
Historically, the scaling laws for the differential cross sections at $s \to \infty$ and fixed $t/s$, where $s$ and $t$ are the Mandelstam variables, were investigated in Refs.~\cite{Brodsky:1973kr,Matveev:1973ra,Lepage:1980fj}.
However, especially for the high energy forward scattering, the cross sections are determined by quite complicated gluonic interactions, and practically it is impossible to calculate those by the perturbative technique of QCD.
Hence, it is required to build a model, which appropriately approximates such a nonperturbative interaction and enables one to predict the cross sections in broad kinematic regions.
It is known that the Reggeon and Pomeron exchange can give a reasonable description for the cross sections.
Donnachie and Landshoff showed that by this combination various hadron-hadron total cross section data can be well described~\cite{Donnachie:1992ny}.
At high energies, the Pomeron exchange, which can be interpreted as a multi-gluon exchange, gives the dominant contribution to the cross sections.
The leading Pomeron trajectory with its intercept $\alpha(0) \approx 1$ but slightly greater than 1, which is the so-called soft Pomeron intercept, can describe the growing behavior of the total hadronic cross sections.

In this work, we consider the Pomeron exchange in the framework of holographic QCD~\cite{Kruczenski:2003be,Son:2003et,Kruczenski:2003uq,Sakai:2004cn,Erlich:2005qh,Sakai:2005yt,DaRold:2005zs,Brodsky:2014yha} to investigate the pion involved scattering processes.
The holographic QCD models are constructed based on the anti-de Sitter/conformal field theory (AdS/CFT) correspondence~\cite{Maldacena:1997re,Gubser:1998bc,Witten:1998qj}, and have been applied to a lot of studies of high energy scattering processes so far~\cite{Polchinski:2001tt,Polchinski:2002jw,BoschiFilho:2005yh,Brower:2006ea,Hatta:2007he,BallonBayona:2007rs,Cornalba:2008sp,Pire:2008zf,Domokos:2009hm,Marquet:2010sf,Cornalba:2010vk,Brower:2010wf,Watanabe:2012uc,Stoffers:2012zw,Watanabe:2013spa,Agozzino:2013zgy,Watanabe:2015mia,Watanabe:2018owy,Xie:2019soz,Burikham:2019zbo,Watanabe:2019zny,FolcoCapossoli:2020pks}.
We extend the work presented in Ref.~\cite{Xie:2019soz}, in which the elastic proton-proton scattering was studied in a holographic QCD model, and investigate the pion-proton and pion-pion scattering, focusing on the Regge regime.
In our model, the Pomeron is described by the Reggeized spin-2 particle propagator, and the proton-Pomeron and pion-Pomeron couplings are given by gravitational form factors of the proton and pion, which can be obtained from the bottom-up AdS/QCD models~\cite{Abidin:2008hn,Abidin:2009hr}.
Combining the propagator and the form factors, we obtain the expressions for the total and differential cross sections, and numerically evaluate them.

There are four parameters in our model, but for three of them, one is the proton-Pomeron coupling constant and two characterize the Pomeron trajectory, we use the values determined in Ref.~\cite{Xie:2019soz} by virtue of the universality of the Pomeron.
Hence, the pion-proton cross sections can be expressed with a single adjustable parameter, which controls the magnitudes of them.
We determine this parameter by the experimental data of the pion-proton total cross section, and then its differential cross section and the pion-pion cross sections can be calculated without any additional parameter.
Although the currently available data are limited, it is found that our calculation is consistent with those.
We will explicitly show our predictions, and discuss the differences between the pion involved processes and the proton-proton case.
The model presented here can be tested at the future experimental facilities.

This paper is organized as follows.
In the next section, we introduce the formalism of our model to describe the high energy pion-proton and pion-pion scattering in the Regge regime.
We explain how to obtain the gravitational form factors, which specify the hadron-Pomeron couplings, from the bottom-up AdS/QCD models in Sec.~\ref{sec:form_factors}.
Our numerical results for the total and differential cross sections are presented in Sec.~\ref{sec:numerical_results}, and the summary of this work is given in Sec.~\ref{sec:summary}.

\section{\label{sec:holographic_model}Holographic description of hadron-hadron scattering in the Regge regime}
The formalism of holographic description of the proton-proton scattering in the Regge regime has been developed in the preceding study~\cite{Domokos:2009hm}.
Here we extend it to the pion-proton and pion-pion scattering and give the expressions for the cross sections.
According to the formalism, the $2^{++}$ glueball field is expressed as a second-rank symmetric traceless tensor $h_{\mu\nu}$, which is assumed to be coupled predominantly to the QCD energy-momentum tensor $T_{\mu\nu}$,
\begin{equation}\label{eq:glueball}
S  = \lambda \int d^4 x h_{\mu \nu} T^{\mu \nu},
\end{equation}
where $\lambda$ is the coupling constant.
Then the hadron-glueball-hadron vertex can be extracted from the matrix element of the energy-momentum tensor between the hadron states,
\begin{equation}\label{eq:element}
\langle
p', s'| T_{\mu \nu} (0) | p, s  %
\rangle.
\end{equation}
Considering the symmetry and conservation of $T_{\mu \nu}$, Eq.~\eqref{eq:element} can be generally expressed in terms of gravitational form factors.

For the proton case, the matrix element can be expressed with three gravitational form factors~\cite{Pagels:1966zza}:
\begin{equation} \label{eq:threeff}
\begin{split}
\langle p', s'| T_{\mu \nu} (0) | p, s \rangle = 
\bar u(p', s') \biggl[ &A_p (t) \frac{ \gamma_\mu  P_\nu + \gamma_\nu  P_\mu }{2}  \\  
+ & B_p(t) \frac{i ( P_\mu \sigma_{\nu \rho} +  P_\nu \sigma_{\mu \rho}) k^\rho}{4m_p} \\  
 +  &C_p(t) \frac{(k_\mu k_\nu - \eta_{\mu \nu} k^2)}{m_p}   \biggr] u(p, s), 
\end{split}
\end{equation}
where $m_p$ represents the proton mass,
$k = p' - p$, %
$t = k^2$, %
and %
$P = (p + p') / 2$.

For the pion case, the matrix element of the energy-momentum tensor can be written in terms of two gravitational form factors:
\ba
&&\left<\pi^a(p_2)|{T}^{\mu\nu}(0)|\pi^b(p_1)\right>=
 \nonumber \\
&&\quad \delta^{ab} \bigg[2 A_\pi(t) p^\mu p^\nu+
\frac{1}{2}C_\pi(t)\left(k^2\eta^{\mu\nu}-k^\mu k^\nu\right)\bigg],
\label{pionMatrix}
\ea
where $p=(p_1+p_2)/2$, $k=p_1 - p_3$, and $t = k^2$.

In this study, we focus on the Regge regime, in which the condition $s \gg |t|$ is satisfied.
In the regime, the contributions from terms which involve the form factors, $B_p$ and $C_p$, can be neglected~\cite{Domokos:2009hm}.
Also, it was shown in Ref.~\cite{Abidin:2008hn} that $C_\pi$ can be expressed with $A_\pi$.
Hence, it is enough to only consider $A_p$ and $A_\pi$ in this study.

The scattering amplitude for the interested hadron-hadron scattering is obtained by combining the Pomeron propagator, which is described by the Reggeized spin-2 particle propagator, and the gravitational form factors of the involved hadrons.
Only $t$-channel scattering is needed to be considered, since it is the dominant channel in the Regge regime.
The lowest state on the leading Pomeron trajectory is assumed to be the $2^{++}$ glueball whose propagator can be written as~\cite{Yamada:1982dx}
\begin{equation} \label{eq:propagator}
	\frac{d_{\alpha\beta\gamma\delta}(k)} 
	{k^2 - m_g^2},
\end{equation}
where $\alpha$ and $\beta$ are the Lorentz indices contracted at one side, and $\gamma$ and $\delta$ are the Lorentz indices contracted at the other side.
$m_g$ is the glueball mass and $d_{\alpha \beta \gamma \delta}$ is explicitly expressed as
\begin{equation}
\begin{split}
	d_{\alpha \beta \gamma \delta} &=
	\frac{1}{2} (\eta_{\alpha \gamma} \eta_{\beta \delta}
	+ \eta_{\alpha \delta} \eta_{\beta \gamma}) \\ %
       & - \frac{1}{2m_g^2}(k_{\alpha} k_{\delta} \eta_{\beta \gamma}
	+ k_{\alpha} k_{\gamma}\eta_{\beta \delta} +
	k_{\beta} k_{\delta} \eta_{\alpha \gamma}
	+ k_{\beta} k_{\gamma} \eta_{\alpha \delta}) \\ %
       & + \frac{1}{24}\left[\left( \frac{k^2}{m_g^2} \right)^2 -
       3\left(\frac{k^2}{m_g^2} \right)
       - 6\right] \eta_{\alpha \beta} \eta_{\gamma \delta} \\%
       &  - \frac{k^2 - 3 m_g^2}{6 m_g^4}(k_{\alpha}k_{\beta}\eta_{\gamma \delta}
       + k_{\gamma} k_{\delta} \eta_{\alpha \beta}) \\ %
      &+ \frac{2k_{\alpha}k_{\beta}k_{\gamma}k_{\delta}}{3 m_g^4}.
\end{split}
\end{equation}
Since we will take the Regge limit in calculation of the scattering amplitude, we can drop the terms which are suppressed by the factor of $t/s$.
Hence only considering the terms in the first line of this equation is actually enough in this study.

The above propagator only includes the exchange of the lightest state, i.e., the $2^{++}$ glueball on the trajectory.
In order to include the higher spin states, the Reggeizing procedure is needed to obtain the full Pomeron propagator.
Following the preceding study~\cite{Domokos:2009hm}, the Reggeized propagator can be obtained by replacing the factor $\frac{1}{k^2 - m_g^2}$ in the glueball propagator with the following one,
\begin{equation}
\frac{- \alpha_c'}{2}\frac{\Gamma [-\chi] \Gamma \left[1 - %
\frac{\alpha_c (t)}{2}\right]}{\Gamma \left[\frac{\alpha_c (t)}{2} 
- \chi -1 \right]} e^{- i \pi \alpha_c (t) / 2}\left(\frac{\alpha'_c s}{2} \right)^{\alpha_c (t) - 2},
\end{equation}
where $ \chi $ is defined by 
\ba \label{chi}
\chi =&&  a_c (s) + a_c (t) + a_c (u) \nonumber \\
       =&& [a_c(0)+a'_cs]+[a_c(0)+a'_ct]+[a_c(0)+a'_cu] \nonumber \\
       =&&3 a_c (0)+a'_c \sum_{i=1}^4 m^2_i,
\ea
with $m_i$ the masses of the scattered particles.
$a_c(x)=a_c(0) + a_c'x$ is a linear function related to the spectrum of the closed strings.
The parameters, $a_c (0)$ and $a'_c$, are related to the parameters, which characterize the Pomeron trajectory, through
\begin{equation}
2 a_c (0) + 2 = \alpha_c (0), \ \ \ 2 a_c' = \alpha_c',
\end{equation}
where $\alpha_c (0)$ and $\alpha_c'$ are the Pomeron intercept and slope, respectively.
With these parameters, $\chi$ in Eq.~\eqref{chi} can be expressed as
\be
\chi=\frac{3}{2} \alpha_c(0) + \frac{1}{2}\alpha_c' \sum_{i=1}^4 m^2_i -3.
\ee
%

\subsection{Pion-proton cross sections}
\begin{figure}[t]
\begin{center}
\includegraphics[width=0.25\textwidth]{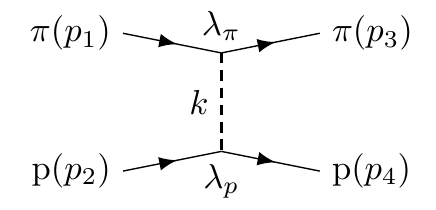}
\caption{\label{fig:diagram1}
The pion-proton scattering with the Pomeron exchange in the $t$-channel.
}
\end{center}
\end{figure}

Here we focus on the pion-proton scattering, which is illustrated in Fig.~\ref{fig:diagram1},
and present the expressions for the differential and total cross sections.
Combining the Pomeron propagator and the pion and proton gravitational form factors, $A_{\pi}(t)$ and $A_{p}(t)$, the amplitude of pion-proton scattering with the Pomeron exchange in the $t$-channel can be expressed as
\ba \label{eq:pi-p}
&&\mathcal{M}_{\pi p} (s, t)=
\lambda_\pi \lambda_p A_{\pi} (t) A_{p} (t) 
\left[ p_1^{\alpha} (\bar{u}_4 \gamma_{\alpha} u_2)  \right]   \nonumber \\ %
&&\times   e^{- i \pi \alpha_c (t) / 2} \Biggl( %
\frac{\Gamma[-\chi]\Gamma\left[1 - %
\frac{\alpha_c (t)}{2}\right]}{\Gamma\left[\frac{\alpha_c (t)}{2} 
- \chi -1 \right]} \Biggr) \left(\frac{\alpha'_c s}{2} \right)^{\alpha_c (t) - 1}, \nonumber \\
\ea
where $\lambda_\pi$ and $\lambda_p$ are the pion-Pomeron and proton-Pomeron coupling constants, respectively.
The parameter $\chi=\frac{3}{2} \alpha_c(0)+\alpha_c'(m^2_{\pi}+m^2_p)-3$, where $m_\pi$ is the pion mass.
Since $p_4 \approx p_2$ in the forward limit and the hadron masses are much smaller than $\sqrt{s}$, taking the average of spin of the incoming proton and summing over spin of the outgoing proton, one finds
\ba
|{\mathcal M}_{\pi p} (s, t)|^2=&&\frac{1}{2} \sum_{\rm 
spin}(\mathcal M \times \mathcal M^*) \nonumber \\ %
=&& \lambda_\pi^2 \lambda_p^2 s^2A^2_{\pi} (t) A^2_{p}(t)  \nonumber \\ %
\times   && \Biggl( %
\frac{\Gamma^2[-\chi]\Gamma^2\left[1 - %
\frac{\alpha_c (t)}{2}\right]}{\Gamma^2 \left[\frac{\alpha_c (t)}{2} 
- \chi -1 \right]} \Biggr) \left(\frac{\alpha'_c s}{2} \right)^{2\alpha_c (t) - 2}. \nonumber \\
\ea
Then the differential cross section of pion-proton scattering is expressed as
\ba
\label{eq:pi-p-DCS}
&&\frac{d \sigma}{dt}\Big|_{\pi p} = \frac{1}{16 \pi s^2}| 
{\mathcal M}_{\pi p} (s,t)|^2 \nonumber \\ %
&&= \frac{\lambda_\pi^2 \lambda_p^2 A^2_{\pi}(t) A^2_{p}(t) \Gamma^2[- \chi] \Gamma^2 \left[1 - %
\frac{\alpha_c (t)}{2}\right]}{16 \pi \Gamma^2 \left[\frac{\alpha_c (t)}{2}  - \chi -1 \right]}
\left(\frac{\alpha'_c s}{2}\right)^{2 \alpha_c (t) - 2}. \nonumber \\
\ea
We apply the optical theorem to Eq.~\eqref{eq:pi-p} and obtain the expression for the total cross section as
\ba\label{eq:pi-p-total}
\sigma_\mathrm{tot}^{\pi p} =&& \frac{1}{s} \ {\rm Im} \ \mathcal{M}_{\pi p} (s, 0) \nonumber \\ %
=&&\frac{\pi \lambda_\pi \lambda_p \Gamma [- \chi]}{\Gamma %
\left[\frac{\alpha_c (0)}{2} \right]\Gamma %
\left[\frac{\alpha_c (0)}{2} - \chi -1 \right]} %
\left(\frac{\alpha_c' s}{2} \right)^{\alpha_c (0) - 1}. \nonumber \\
\ea
%

\subsection{Pion-pion cross sections}
\begin{figure}[t]
\begin{center}
\includegraphics[width=0.25\textwidth]{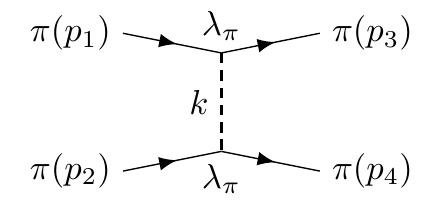}
\caption{\label{fig:diagram2}
The pion-pion scattering with the Pomeron exchange in the $t$-channel.
}
\end{center}
\end{figure}

The pion-pion scattering is illustrated in Fig.~\ref{fig:diagram2}.
Combining the Pomeron propagator and the pion gravitational form factor $A_{\pi}(t)$, the scattering amplitude can be written as
\ba
\mathcal{M}_{\pi \pi} (s, t)=&&
\lambda_\pi^2 s A_{\pi}^2 (t) 
e^{- i \pi \alpha_c (t) / 2} \nonumber \\ %
&& \times \Biggl( %
\frac{\Gamma[-\chi]\Gamma\left[1 - %
\frac{\alpha_c (t)}{2}\right]}{\Gamma\left[\frac{\alpha_c (t)}{2} 
- \chi -1 \right]} \Biggr) \left(\frac{\alpha'_c s}{2} \right)^{\alpha_c (t) - 1}, \nonumber \\
\ea
where $\chi=\frac{3}{2} \alpha_c(0)+2\alpha_c'  m^2_{\pi} -3$.
Then the differential cross section is expressed as
\ba
\label{eq:pi-pi-DCS}
\frac{d \sigma}{dt}\Big|_{\pi \pi}=&&\frac{1}{16 \pi s^2}| 
\mathcal{M}_{\pi \pi} (s,t)|^2 \nonumber \\ %
=&&  \frac{\lambda_\pi^4 A_{\pi}^4 (t) \Gamma^2[- \chi] \Gamma^2 \left[1 - %
\frac{\alpha_c (t)}{2}\right]}{16 \pi \Gamma^2 \left[\frac{\alpha_c (t)}{2} - \chi -1 \right]}
\left(\frac{\alpha'_c s}{2}\right)^{2 \alpha_c (t) - 2}. \nonumber \\
\ea
The total cross section is given by
\ba\label{eq:pi-pi-total}
\sigma_\mathrm{tot}^{\pi \pi} = && \frac{1}{s} \ {\rm Im} \ \mathcal{M}_{\pi \pi} (s, 0) \nonumber \\ %
=&&\frac{\pi \lambda_\pi^2 \Gamma [- \chi]}{\Gamma %
\left[\frac{\alpha_c (0)}{2} \right]\Gamma %
\left[\frac{\alpha_c (0)}{2} - \chi -1 \right]} %
\left(\frac{\alpha_c' s}{2} \right)^{\alpha_c (0) - 1}. \nonumber \\
\ea
%

\section{\label{sec:form_factors}Gravitational form factors of the pion and proton}
To numerically evaluate the differential cross sections presented in the previous section, we need to specify the gravitational form factors, $A_p$ and $A_\pi$, which can be obtained from the bottom-up AdS/QCD models~\cite{Abidin:2008hn,Abidin:2009hr} in the five-dimentional AdS space with the metric,
\be
ds^2= g_{MN} dx^M dx^N %
= \frac{1}{z^2} \eta_{MN} dx^M dx^N,	\quad \varepsilon<z<z_0,
\ee
where $\eta_{MN}=\text{diag}(1,-1,-1,-1,-1)$.
In this study, we adopt the hard-wall model, in which the AdS geometry is sharply cut off in the infrared region at $z=z_0$ to break the conformal symmetry and introduce the QCD scale.

According to the AdS/CFT correspondence, the energy-momentum tensor operator $T_{\mu \nu}$ in the four-dimensional strongly coupled theory corresponds to a field in the five-dimensional AdS space:
\ba
T_{\mu\nu}(x)  &\leftrightarrow& h_{\mu\nu}(x,z) ,
\ea
where $h_{\mu\nu}$ represents variations of the metric tensor,
\be
g_{\mu\nu}(x,z) = \frac{1}{z^2} \left( \eta_{\mu\nu} + h_{\mu\nu}(x,z) \right). \label{AdSmetric}
\ee
Considering the $h \psi \psi$ terms, where $\psi$ represents the hadron field, in the classical action of the bottom-up AdS/QCD model, the gravitational form factors can be obtained.

\subsection{Pion gravitational form factor}
The effective action for the meson fields is given by~\cite{Erlich:2005qh}
\begin{align}
S_{M}  = \int {d^5 x} \sqrt g \biggl[ \mathrm{Tr} \biggl\{ &\left| {DX} \right|^2  + 3\left| X \right|^2 \nonumber \\ %
&- \frac{1}{{4g_5^2 }}\left( {F_L^2  + F_R^2 } \right) \biggr\} \biggr] ,
\end{align}
where %
the covariant derivative $D^M X = \partial^M X - i A_L^M X + i X A_R^M$ is introduced with the gauge fields, $A_L^M$ and $A_R^M$, %
and the field strength tensor $F^{MN}_{L,R} = \partial^M A^N_{L,R} - \partial^N A^M_{L,R} - i [A^M_{L,R}, A^N_{L,R}]$. %
The bulk field $X$ is defined by $X(x,z) = X_0(z) \exp (2i t^a \pi^a)$ with the bulk scalar $X_0$, the isospin operator $t^a$, and the pion field $\pi^a$.
The bulk scalar is given by
\begin{align}
X_0 = \frac{1}{2}\mathbb{I} v(z) %
= \frac{1}{2}\mathbb{I} \left( m_q z+\sigma z^3 \right),
\end{align}
where $m_q$ and $\sigma$ can be identified as the quark mass and the chiral condensate, respectively.
Since we take the chiral limit for simplicity in this study, $m_q = 0$ and the pion is massless.

The action containing the pion field $\pi$ and the axial-vector field $A=(A_L-A_R)/2$ up to the second order is expressed as
\ba
S_A &=& \int d^5 x \sqrt{g} \,  \bigg[ \frac{v(z)^2}{2} g^{MN}
(\partial_M\pi^a-A^a_M) (\partial_N\pi^a-A^a_N)
\nonumber \\ %
&& 	\hskip 17 mm	
-~  \frac{1}{4g_5^2} g^{KL} g^{MN}  {F^a}_{KM}F^a_{LN}   \bigg].\label{axialAction}
\ea
Taking the variation over $A_M^a$ of this equation, the equations of motion are obtained, and the pion wave function is given by~\cite{Grigoryan:2007wn}
\ba
\Psi(z)=&&z\Gamma \left[ \frac{2}{3} \right]\left(\frac{\alpha}{2}\right)^{\frac{1}{3}} \nonumber \\ %
&&\times\left(I_{-\frac{1}{3}}(\alpha z^3)-I_{\frac{1}{3}}(\alpha z^3) 
\frac{I_{\frac{2}{3}} \left( \alpha (z_0^\pi)^3 \right)}{I_{-\frac{2}{3}} \left( \alpha (z_0^\pi)^3 \right)}\right),
\ea
where $\alpha= g_5\sigma/3$ with $g_5=2\pi$.
The cutoff parameter $z_0^\pi$ is determined with the $\rho$-meson mass $m_\rho$ by the condition, %
$J_0 (m_\rho z_0^\pi) = 0$. %
The obtained value is $z_0^\pi = 1/(322~\mathrm{MeV})$.
The parameter $\sigma$ is related to the pion decay constant:
\be
f^2_\pi=\frac{3}{4\pi^2}\frac{\Gamma(2/3)}{\Gamma(1/3)}\left(2\alpha^2\right)^{\frac{1}{3}} %
\frac{I_{\frac{2}{3}}\left( \alpha (z_0^\pi)^3 \right)}{I_{-\frac{2}{3}}\left( \alpha (z_0^\pi)^3 \right)}.
\ee
Using the experimental value $f_\pi = 92.4~\mathrm{MeV}$, it is found to be $\sigma = (332~\mathrm{MeV})^3$.

\begin{figure}[t]
\begin{center}
\includegraphics[width=0.45\textwidth]{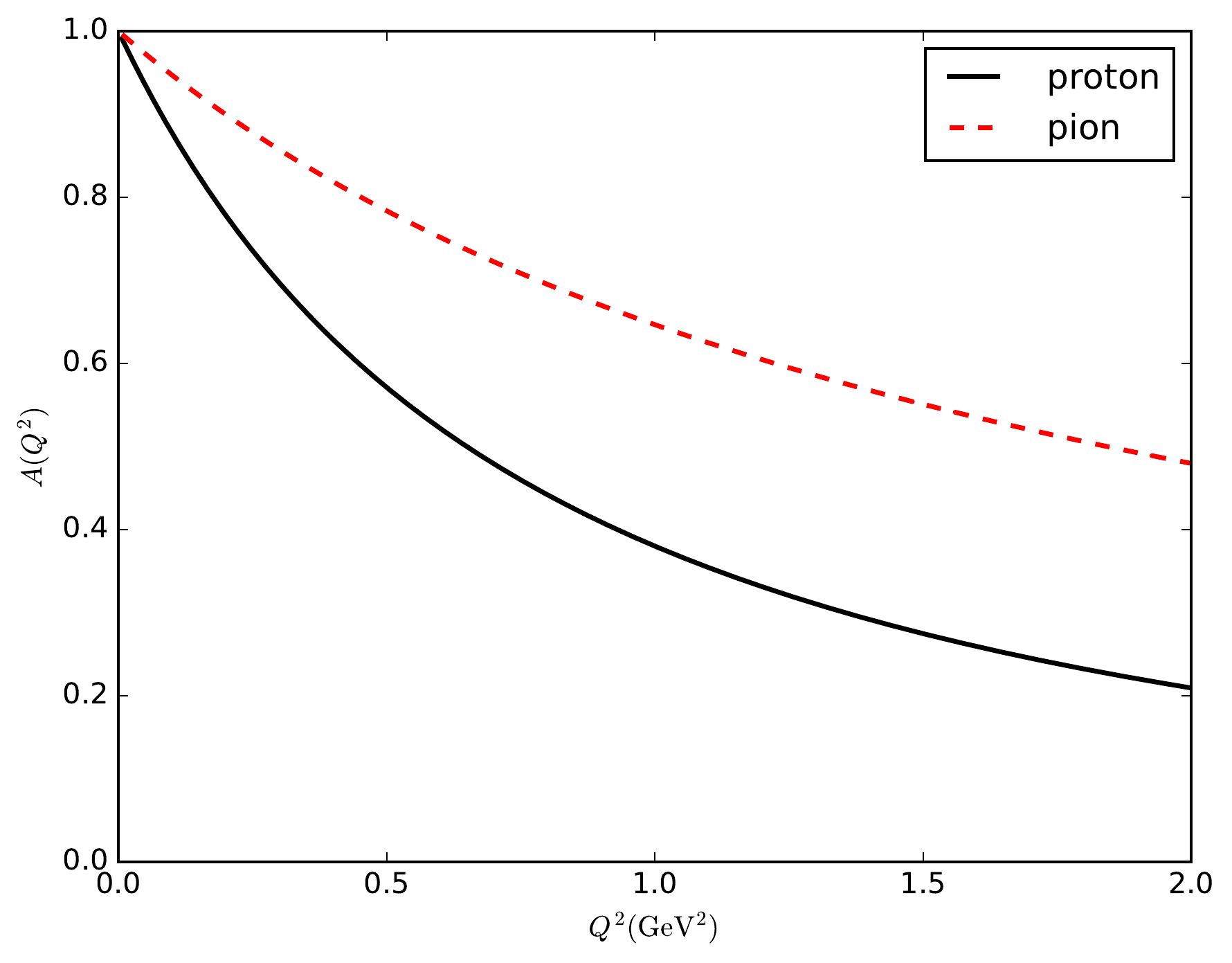}
\caption{\label{fig:form_factors}
The gravitational form factors as a function of $Q^2$.
The solid and dashed curves represent the proton and pion results, respectively.
}
\end{center}
\end{figure}

To obtain the matrix element of the energy-momentum tensor, we need to consider the three-point function,
\ba
&& \big< 0 \big| {\mathcal T} {J}_5^{a\alpha}(x) \hat{T}_{\mu\nu}(y){J}_5^{b\beta}(w)\big| 0 \big>	 \nonumber \\ %
&& \hskip 19 mm= -
 \frac{    2 \, \delta^3 S}{\delta A^{a0}_\alpha (x) \delta h^{\mu\nu 0}(y) \delta A^{b0}_\beta(w)} .
\ea
Focusing on the relevant part of the action, the matrix element is expressed as
\ba
  &&\left<\pi^a(p_2)\right|\hat{T}^{\mu\nu}(0)\left|\pi^b(p_1)\right>
  \nonumber \\ %
  &&\ \ \ = 2 \delta^{ab} A_\pi(Q^2)\bigg[ p^\mu p^\nu+\frac{1}{12}
  \left(q^2\eta^{\mu\nu}-q^\mu q^\nu\right)\bigg]. \label{pionTracelessMatrix}
\ea
The gravitational form factor is given by
\be
 A_\pi(Q) = \int dz\, \mathcal{H} (Q,z) \left(\frac{(\partial_z \Psi(z))^2}{g_5^2f_\pi^2z} + \frac{v(z)^2\Psi(z)^2}{f_\pi^2 z^3}\right),  \label{eq:ff_pion}
\ee
where $\mathcal{H}(Q,z)$ is the bulk-to-boundary propagator of the graviton for the spacelike momentum transfer, and related to the metric perturbation.
As the solution of the linearized Einstein equation, it is given by~\cite{Abidin:2008ku}
\be
\mathcal{H}(Q,z) = \frac{1}{2}Q^2 z^2\bigg(\frac{K_1(Q z_0)}{I_1(Q z_0)}I_2(Qz) + K_2(Qz)\bigg). \label{H}
\ee
With these expressions, the $Q^2$ dependence of the pion gravitational form factor is shown in Fig.~\ref{fig:form_factors}.
%

\subsection{Proton gravitational form factor}
Similar to the pion case, the proton gravitational form factor can also be obtained from the bottom-up AdS/QCD model~\cite{Abidin:2009hr}, in which the proton is described by a solution of the five-dimensional Dirac equation.
The classical actions is given by~\cite{Hong:2006ta}
\ba
  \label{eq:proton_action}
S_F = %
  \int d^5 x && \sqrt{g} e^{- \Phi (z)} \bigg( \frac{i}{2} \bar{\Psi} e^N_A \Gamma^A D_N \Psi \nonumber \\ %
  &&- \frac{i}{2} (D_N \Psi)^\dagger \Gamma^0 e^N_A \Gamma^A \Psi - M \bar{\Psi}\Psi \bigg),
\ea
where %
$e^N_A = z \delta^N_A$, %
$D_N = \partial_N + \frac{1}{8} \omega_{NAB}[\Gamma^A, \Gamma^B] - i V_N$ is the covariant derivative,
and $M$ represents the mass of the bulk spinor.

The field $\Psi$ is obtained as a solution of the Dirac equation.
Defining the right-handed and left-handed spinor $\Psi_{R,L} = (1/2)(1\pm \gamma^5)\Psi$ and imposing appropriate boundary conditions, the normalizable modes are given by
\ba
\psi_L^{(n)} (z) =&& \frac{\sqrt{2} z^\beta J_\beta (m_n^p z)}{z_0 J_\beta (m_n^p z_0^p)}, \nonumber \\ %
\psi_R^{(n)} (z) =&& \frac{\sqrt{2} z^\beta J_{\beta - 1} (m_n^p z)}{z_0 J_\beta (m_n^p z_0^p)},
\ea
where $\beta = M + \frac{1}{2}$ and $m^p_n$ is the mass of the $n$-th Kaluza-Klein state, which satisfies the condition $J_{\beta-1}(m_n^p z_0^p)=0$.
Since it can be seen in Ref.~\cite{Abidin:2009hr} that $M=\frac{3}{2}$ realizes the correct large momentum scaling for the proton electromagnetic form factor and we only consider the ground state of the proton in this study, following this condition the cutoff parameter is determined with the proton mass and found to be $z_0^p = 1/(245~\mathrm{MeV})$.

The matrix element of the energy-momentum tensor can be extracted from the three-point function,
\begin{equation} \label{eq:3-point}
\left< 0 \big| \mathcal{T} \mathcal{O}^i_R (x) %
T^{\mu \nu} (y) \bar{ \mathcal{O}}^j_R (w) \big| 0 \right>.
\end{equation}
Focusing on the relevant part of the action, which has the form of $h \bar{\Psi} \Psi$, and comparing the Lorentz structure with that of Eq.~\eqref{eq:threeff}, the proton gravitational form factor is given by
\begin{equation}
A_p (Q) = %
\int dz \frac{1}{2 z^{2 M}} \mathcal{H} (Q, z) %
\left(\psi_L^2 (z) + \psi_R^2 (z) \right).
\end{equation}
The $Q^2$ dependence of $A_p$ is shown in Fig.~\ref{fig:form_factors}, from which it is seen that $A_p$ decreases substantially faster than $A_\pi$ with $Q^2$.

\section{\label{sec:numerical_results}Numerical results}
The numerical results obtained from the expressions explained in the previous sections are presented here.
In the present model, there are four parameters in total, $\alpha_c(0)$, $\alpha_c'$, $\lambda_\pi$, and $\lambda_p$, which need to be determined by the experimental data.
However, the three of those were determined in the preceding study of the proton-proton scattering~\cite{Xie:2019soz}, and we can use the resulting values for the pion-proton and pion-pion cases by virtue of the universality of the Pomeron.
In Ref.~\cite{Xie:2019soz}, the parameters were determined with all the available data, including the recent TOTEM ones~\cite{TOTEM:2013lle,TOTEM:2016lxj,TOTEM:2017asr,TOTEM:2017sdy,TOTEM:2018hki,TOTEM:2018psk}, for the total and differential cross sections at high energies, and it was shown that the model can well describe the data.
The parameter values taken from that work are $\alpha_c(0)=1.086$, $\alpha_c'=0.377$~GeV$^{-2}$, and $\lambda_p=9.70$~GeV$^{-1}$.
Hence, the total and differential cross sections of pion-proton and pion-pion scattering are now expressed with a single adjustable parameter $\lambda_\pi$ which describes the pion-Pomeron coupling.

We determine the parameter $\lambda_\pi$ by the pion-proton total cross section data summarized by the Particle Data Group in 2020~\cite{ParticleDataGroup:2020ssz}.
Since we focus on the Regge regime and only consider the Pomeron exchange in this study, we need to set a kinematic cut to suppress the contributions from the Reggeon exchange.
We utilize all the available data that have $P_{\mathrm{lab}} \ge 100~\mathrm{GeV}/c$, which corresponds to $\sqrt{s} \ge$~13.73~GeV, to determine the value of $\lambda_\pi$.
Using Eq.~\eqref{eq:pi-p-total} and the MINUIT package~\cite{James:1975dr}, the numerical fit is performed, and the best fit value is found to be $\lambda_\pi=3.950 \pm 0.016$~GeV$^{-1}$.

\begin{figure}[t]
\begin{center}
\includegraphics[width=0.45\textwidth]{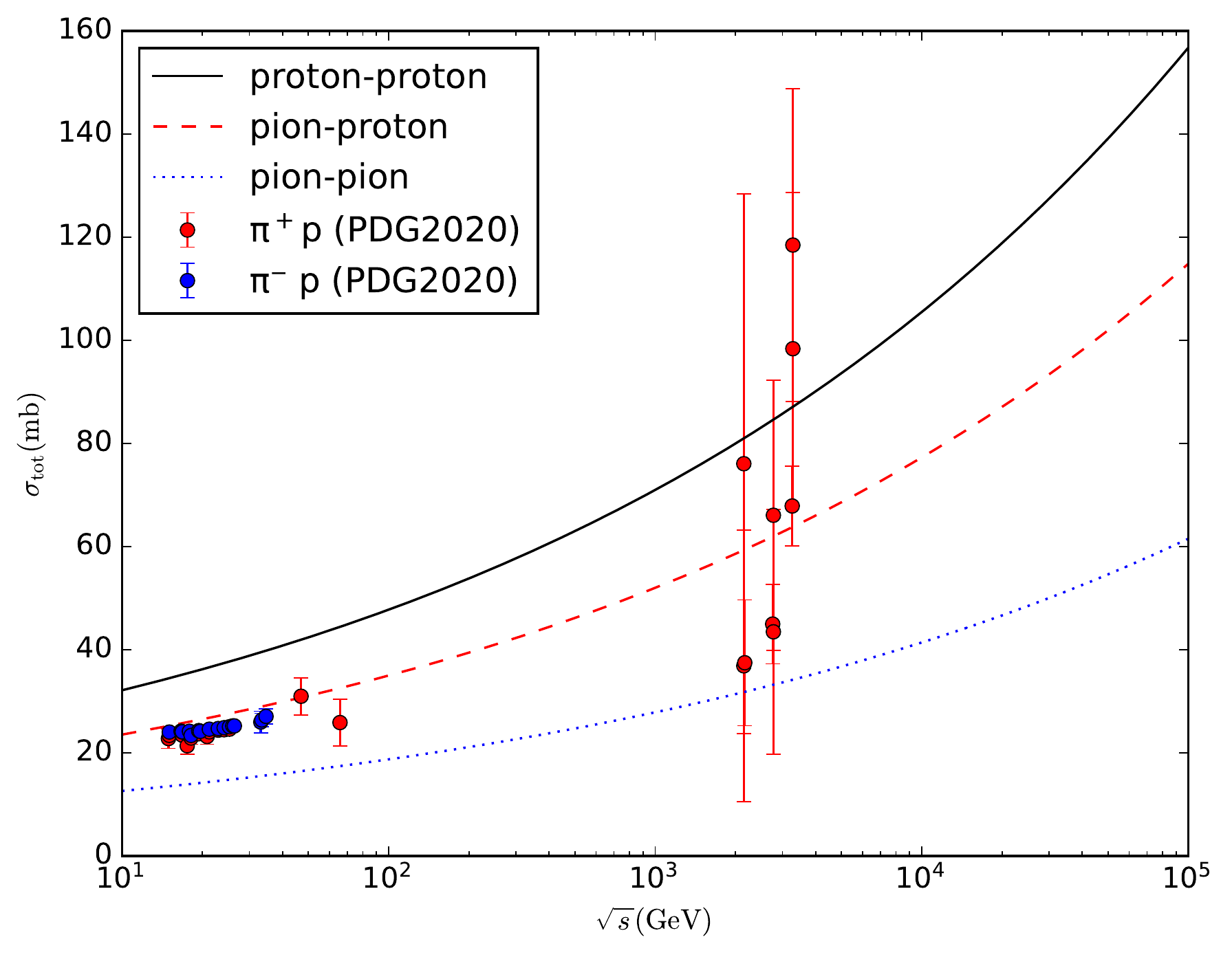}
\caption{\label{fig:TCS}
The total cross sections as a function of $\sqrt{s}$.
The solid, dashed, and dotted curves represent the results for proton-proton~\cite{Xie:2019soz}, pion-proton, and pion-pion scattering, respectively.
The experimental data summarized by the Particle Data Group in 2020~\cite{ParticleDataGroup:2020ssz} are depicted by circles with error bars.
}
\end{center}
\end{figure}

Once the parameter $\lambda_\pi$ is determined, the pion-pion total cross section can be calculated without any additional parameter, using Eq.~\eqref{eq:pi-pi-total}.
The resulting total cross sections are shown in Fig.~\ref{fig:TCS},
in which the proton-proton result taken from Ref.~\cite{Xie:2019soz} is also shown for comparison.
Although the number of available experimental data in the considered kinematic region is limited and the data in the TeV scale have huge uncertainties, it can be seen from the figure that our calculation is consistent with those.
In the present model, since the parameters which characterize the Pomeron trajectory are common to all the hadron-hadron processes, the total cross section rations are constant.
The resulting ratios are found to be
\ba
\frac{\sigma_{\mathrm{tot}}^{\pi p}}{\sigma_{\mathrm{tot}}^{p p}} = 0.73,
\ \ \
\frac{\sigma_{\mathrm{tot}}^{\pi \pi}}{\sigma_{\mathrm{tot}}^{p p}} = 0.39.
\ea

\begin{figure*}[t]
\begin{center}
\includegraphics[width=0.9\textwidth]{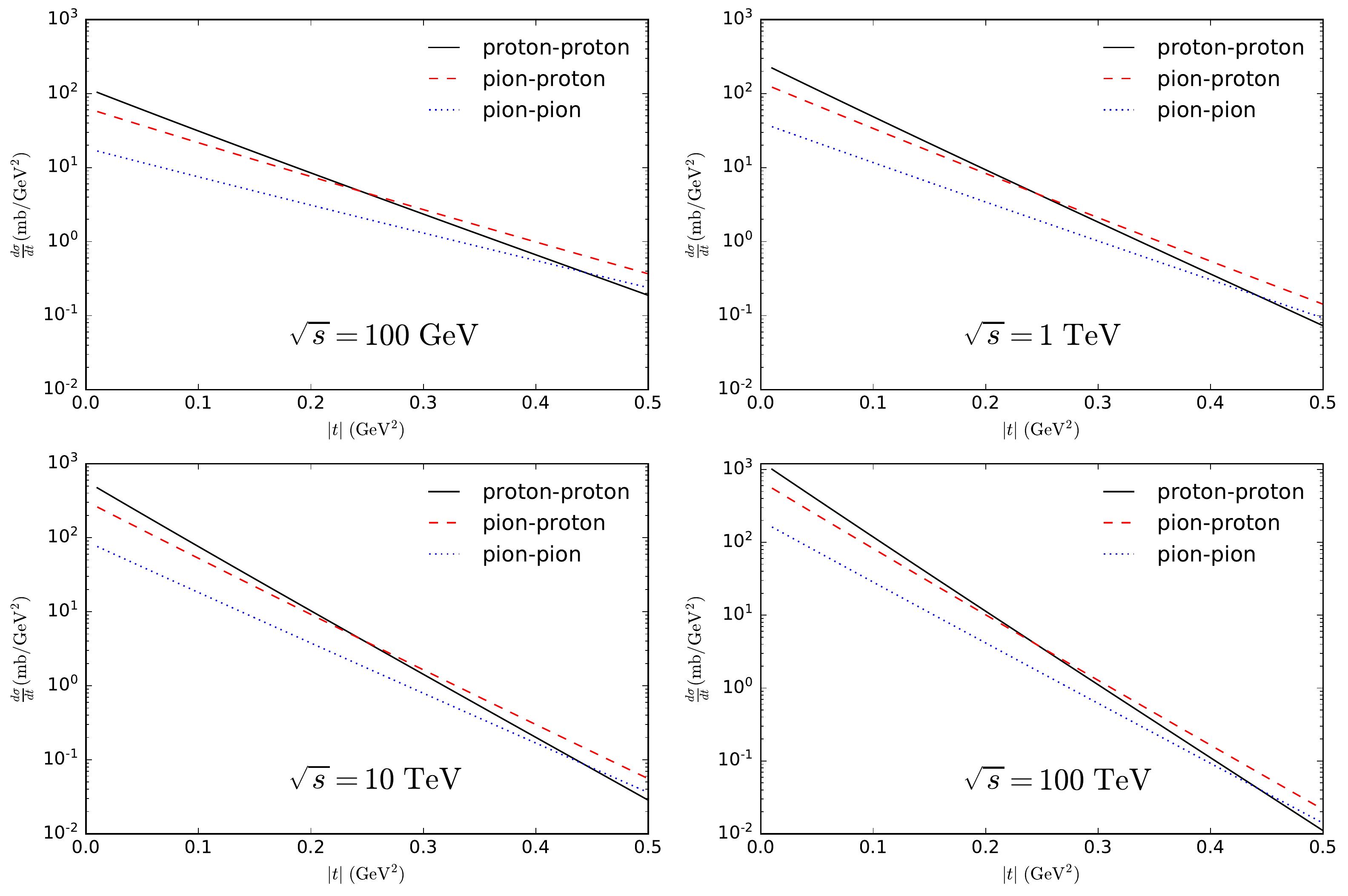}
\caption{\label{fig:DCS}
The differential cross sections as a function of $|t|$ for various $\sqrt{s}$.
The solid, dashed, and dotted lines represent the results for proton-proton, pion-proton, and pion-pion scattering, respectively.
The proton-proton results are obtained from the model setup presented in Ref.~\cite{Xie:2019soz}.
}
\end{center}
\end{figure*}

Using Eqs.~\eqref{eq:pi-p-DCS} and \eqref{eq:pi-pi-DCS}, and the gravitational form factors specified in Sec.~\ref{sec:form_factors}, the pion-proton and pion-pion differential cross sections can be numerically evaluated.
The results are shown in Fig.~\ref{fig:DCS},
in which the proton-proton results obtained from the model setup presented in Ref.~\cite{Xie:2019soz} are also shown for comparison.
From the figure, the substantial differences of the $t$ dependence among the three curves can be clearly seen.
Comparing to the proton-proton case, the $t$ dependence of the pion-proton results is obviously weaker, and that of the pion-pion results is further weaker.
This qualitative behavior reflects the substantial $t$ dependence difference between the proton and pion gravitational form factors, and is independent of the $\sqrt{s}$ values in the considered kinematic regime.

\section{\label{sec:summary}Summary and discussion}
We have investigated the elastic pion-proton and pion-pion scattering at high energies in a holographic QCD model, focusing on the Regge regime.
Considering the Pomeron exchange, which is expressed by the Reggeized spin-2 particle propagator, and specifying the hadron-Pomeron couplings by the gravitational form factors, which are obtained from the bottom-up AdS/QCD models, we have proposed a novel description of those scattering processes.
Since this model is overall constructed within the framework of holographic QCD, this is a consistent description.
The notable advantage of the model is the fact that it only includes a single adjustable parameter, since the other three parameters were precisely determined in the preceding study of the proton-proton scattering.

We have determined this single parameter by the experimental data of the pion-proton total cross section and shown that our calculation is consistent with those data.
Then, we have presented our predictions for the pion-pion total cross section, and the pion-proton and pion-pion differential cross sections, which can be calculated without any additional parameter in the present model.
From the differential cross section results, the $t$ dependence differences are clearly seen, which reflect the substantial difference between the proton and pion gravitational form factors.

In this study, we have performed the numerical fit with the pion-proton data, and the resulting pion-proton total cross section seems reasonable.
However, the number of the currently available data is limited and also the data in the TeV scale have huge uncertainties.
It is obvious that the determination of the best fit value was strongly affected by the data in the lower $s$ region.
Hence, the absolute magnitudes of the cross sections presented in this paper may have some uncertainties, although it is difficult to estimate those.
This is our best effort at this moment, and more data especially in the high $s$ region, where the contributions from the Reggeon exchange are completely suppressed, are required to pin down the absolute magnitudes.
Nevertheless, it is interesting that the qualitative differences are found in our resulting differential cross sections.
In the present model, the slope depends on the slope parameter of the Pomeron trajectory and the gravitational form factor, and is especially sensitive to the latter one.
Therefore, it is expected that future experimental data will help to improve our understandings about the pion structure and also QCD itself.

\section*{Acknowledgments}
This work was partially supported by the National Natural Science Foundation of China under Grant Nos. 11875178 and 12005114.
The work of A.W. was supported by the start-up funding from China Three Gorges University.
A.W. is also grateful for the support from the Chutian Scholar Program of Hubei Province.

\bibliography{reference}

\end{document}